\documentclass[twocolumn,prb]{revtex4}%
\usepackage{amsmath}
\usepackage{graphicx}
\usepackage{amsfonts}
\usepackage{amssymb}%
\providecommand{\U}[1]{\protect\rule{.1in}{.1in}}
\providecommand{\U}[1]{\protect\rule{.1in}{.1in}}

\begin{document}
\title{Hot electron relaxation in a heavy fermion system with tuned doping.}
\author{J. Demsar,$^{1,2,3}$ V.V. Kabanov,$^{1,3}$ A.S. Alexandrov,$^{4}$ H.J.
Lee,$^{5}$ E.D. Bauer,$^{5}$ J.L. Sarrao,$^{5}$ A.J. Taylor$^{5}$}
\affiliation{$^{1}$Physics Dept. and Zukunftskolleg, University of Konstanz, D-78457
Konstanz, Germany}
\affiliation{$^{2}$Center for Applied Photonics, University of Konstanz, D-78457
Konstanz, Germany}
\affiliation{$^{3}$Complex Matter Dept., Jozef Stefan Institute, SI-1000, Ljubljana,
Slovenia }
\affiliation{$^{4}$Dept. of Physics, Loughborough University, Loughborough LE11 3TU, UK }
\affiliation{$^{5}$Los Alamos National Laboratory, Los Alamos, NM 87545, USA }
\date{}

\begin{abstract}
Femtosecond time-resolved optical spectroscopy was used to systematically
study photoexcited carrier relaxation dynamics in the intermediate-valence
heavy fermion system Yb$_{1-x}$Lu$_{x}$Al$_{3}$ ($0\leq x\leq1$). Given the
demonstrated sensitivity of this experimental technique to the presence of the
low energy gaps in the charge excitation spectrum, the aim of this work was to
study the effect of dilution of the Kondo lattice on its low energy electronic
structure. The results imply that in Yb$_{1-x}$Lu$_{x}$Al$_{3}$ the
hybridization gap, resulting from hybridization of local moments and
conduction electrons, persists up to 30\% doping. Interestingly, below some
characteristic, doping dependent temperature $T^{\ast}(x)$ the relaxation time
divergence, governed by the relaxation bottleneck due to the presence of the
indirect hybridization gap, is truncated. This observation is attributed to
the competing ballistic transport of hot electrons out of the probed volume at low
temperatures. The derived theoretical model accounts for both the functional
form of relaxation dynamics below $T^{\ast}(x)$, as well as the doping
dependence of the low temperature relaxation rate in Yb$_{1-x}$Lu$_{x}$%
Al$_{3}$.

\end{abstract}
\maketitle

\section{Introduction}

A reasonable understanding of the Kondo problem was achieved in the case of
isolated Kondo impurities by the Anderson impurity model (AIM). Despite the
fact that many heavy electron systems are stoichiometric with a local moment
in each unit cell, many of their physical properties can be well accounted for
by the AIM. On the other hand, some low-temperature properties such as optical
conductivity are fundamentally different than those expected of the AIM. These
are consistent with the renormalized band behavior, captured by the Anderson
lattice model. Here hybridization of local moments and conduction electrons
leads to the presence of the hybridization gap (HG) in the vicinity of the
Fermi level,\cite{Reviews,Degiorgi,Riseborough} where in heavy fermions the
Fermi level lies close to the edge of the lower hybridized
band.\cite{HE2006,Review2006}

Recent experiments on the photoexcited carrier relaxation dynamics in several
heavy electron compounds have shown that the relaxation of the electronic
system back to equilibrium is extremely sensitive to the underlying low energy
electronic structure and strongly depends on the temperature ($T$) and
excitation level.\cite{HE2006,Review2006,HF2003,Ahn2004,Burch2008} It has been
shown, that the bottleneck in carrier relaxation is governed by the presence
of the HG near the Fermi level.\cite{HE2006,Review2006} Both $T$ and
excitation density dependence of the relaxation process can be well accounted
for by the phenomenological Rothwarf-Taylor (RT) model
\cite{Review2006,RT,MgB2,KabanovRT} which was originally developed to describe
the relaxation in superconductors driven out of
equilibrium.\cite{RT,MgB2,KabanovRT} We should note that the alterantive model
presented in Refs. [6,7] is esentially also a bottleneck model. Here the energy
and momentum conservation law leads to the suppression of e-ph scattering near
E$_{f}$ in the case of sound velocity exceeding the Fermi velocity. This
suppression of e-ph scattering near E$_{f}$ effectivly acts as a gap in the
DOS. However, since the temperature and excitation dependence of the
relaxation dynamics in a Kondo insulator SmB$_{6}$ is nearly identical to that
of heavy fermion systems,\cite{HE2006,Review2006} it was naturally to assume
that the origin of the bottleneck in the entire class of materials was in the
presence of the indirect hybridization gap.\cite{HE2006,Review2006}

In this paper we report first systematic studies of photoexcited carrier
relaxation dynamics in an intermediate-valence system Yb$_{1-x}$Lu$_{x}%
$Al$_{3}$. In Yb$_{1-x}$Lu$_{x}$Al$_{3}$ concentration of local f moments can
be continuously varied between 1 (YbAl$_{3}$ is a heavy fermion with the Kondo
temperature of 600-700 K) and 0 by replacing Yb with open f-shell with Lu with
closed f-shell (LuAl$_{3}$ is a normal metal). Dilution by doping with
non-magnetic ions should give rise to the disappearance of the long range
order, and the associated HG at some critical doping. Moreover, in YbAl$_{3}$
the low-temperature anomalies observed in susceptibility and specific heat
suggest the presence of a second energy scale of the order of 40
K.\cite{Cornelius} The observed anomalies are quickly suppressed by doping
\cite{Bauer04} suggesting that this energy scale is related to the onset of
Fermi liquid coherence.\cite{Cornelius,Bauer04} Using optical pump-probe
spectroscopy in the low perturbation regime we show that at high $T$ and
doping levels $0\leq x\leq0.3$ the relaxation rate decreases with decreasing
$T$ much like in other heavy electron systems studied so
far.\cite{HE2006,Review2006,HF2003} This behavior is consistent with the
relaxation bottleneck due to the presence of the HG.\cite{HE2006} The HG does
not change significantly in the doping range $0\leq x\leq0.3$, and the
disappearance of the long range order appears near $x\approx0.4$. 

At some characteristic doping dependent temperature $T^{\ast}(x)$, the slowing
down of relaxation is truncated and the relaxation rate becomes constant below
$T^{\ast}$. This behavior is explained by the competition between the
relaxation across the HG and the ballistic electron transport, when at low $T$
the electron mean free path $l$ becomes larger than the optical penetration
depth $\lambda$. We derived the theoretical model that describes the
functional form of relaxation for such a case and discuss the striking doping
dependence of the low temperature relaxation rate.

\section{Experimental}

The experiments were performed in the standard pump-probe configuration using
a mode-locked Ti:sapphire laser as the source for both photoexcitation and
probe laser pulse trains. The pump and probe pulses had a duration of $<20$ fs
at the center wavelength of 800 nm. The experiments were performed in the very
weak excitation regime with the excitation fluence of $F\approx0.4$ $\mu
$J/cm$^{2}$.\cite{comment2} From the literature values \cite{OptCond} of the
complex conductivity at 1.5 eV  we estimated the corresponding absorbed energy
densities to be $\approx90$ mJ/cm$^{3}$ in YbAl$_{3}$. Based on the known
values of the Sommerfeld constant we estimated the corresponding increase in
the electronic temperature after excitation to be less than 10 K over the
entire $T$ range implying that the dynamics being studied is near the thermal
equilibrium. The Yb$_{1-x}$Lu$_{x}$Al$_{3}$ samples (10 doping levels were
studied) were grown by the \textquotedblleft self-flux\textquotedblright%
\ method in excess Al.\cite{Bauer04}

\section{Results}

Figure 1 shows the dynamics of photoinduced reflectivity change in YbAl$_{3}$
and Yb$_{0.95}$Lu$_{0.05}$Al$_{3}$ in the $T$ range between 4 K and 300 K. At
300 K, the rise time of about 70 fs is followed by fast electronic relaxation
which can be well described by a single exponential decay with the relaxation
time $\tau=330$ fs. Following this initial electronic relaxation the system
recovers to equilibrium on the timescale of several 100 ps, which is
attributed to heat flow out of the excitation volume. Upon lowering $T$,
changes in both the amplitude of the transient, \textsf{A}, and $\tau$ are
observed. Upon cooling $\tau$ gradually increases in the same fashion as in
previously studied heavy electron compounds,\cite{HE2006,Review2006,HF2003}
reaching 3.2 ps at $\approx50$ K. There the anomalous increase in $\tau$ is
truncated and $\tau$ remains constant below $\approx50$ K. The amplitude of
transient, \textsf{A,} also shows a pronounced $T$ dependence similar to early
reports.\cite{HE2006} At low temperatures \textsf{A }is being nearly constant
up to $\approx50$ K and showing a pronounced suppression upon further $T$
increase. Qualitatively similar results are obtained for all $0<x<0.3$, except
that the temperature where the relaxation time divergence is truncated,
$T^{\ast}$, is doping dependent, as discussed below. In LuAl$_{3}$, however,
the dynamics follows the behavior seen in normal metals, where the relaxation
is only very weakly temperature dependent with the relaxation time slightly
increasing upon increasing the temperature.\cite{Groeneveld} Similarly,
virtually no T dependence of $\mathcal{A}$ is observed.

\begin{figure}[h]
\begin{center}
\includegraphics[clip=true,width=1.\linewidth]{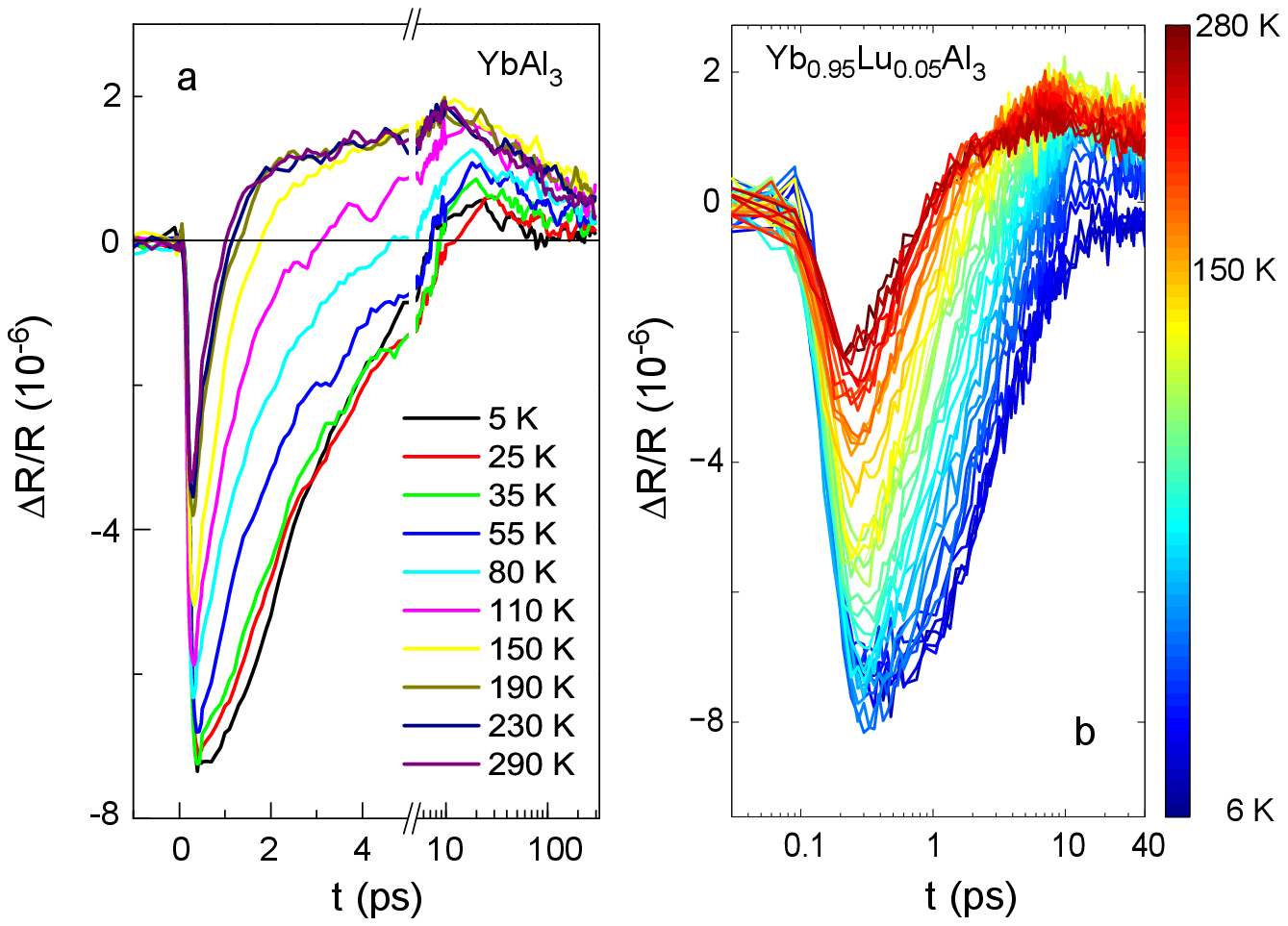}
\end{center}
\caption{(color online) Photoinduced reflectivity traces in a) YbAl$_{3}$ at
selected temperatures and b) in Yb$_{0.95}$Lu$_{0.05}$Al$_{3}$ from 6-280 K
(in 6K steps between 6-140K and 10K steps from 140-280 K).}%
\end{figure}

The $T$ dependence of $\tau$ in YbAl$_{3}$ is shown in Figure 2. Above
$\approx50$ K it is consistent with the relaxation bottleneck scenario due to
the presence of the narrow HG, $E_{hg}$, in the density of
states.\cite{HE2006,Review2006} The relaxation process can be described by the
RT model. Here following photoexcitation and the initial electron-electron and
electron-phonon collisions which proceed on a sub-ps timescale, the system is
characterized by excess densities of electron-hole pairs (EHPs) and high
frequency phonons (HFP). When an EHP with an energy $\geq E_{hg}$ recombines,
a high frequency phonon ($\omega>E_{hg}$) is created. Since HFP can
subsequently excite EHP, the recovery is essentially governed\ by the decay of
the HFP population.\cite{RT,KabanovRT} In the low excitation limit, as in this
case, the $T$ dependencies of both amplitude of the induced reflectivity
change \textsf{A, }and $\tau$ are governed by the $T$ dependence of the number
density of thermally excited EHPs, $n_{T}$. It was shown \cite{KabanovRT} that
$n_{T}\propto\mathcal{A}^{-1}-1$, where $\mathcal{A}(T)=$ \textsf{A}%
$(T)/$\textsf{A}$(T\longrightarrow0)$ and
\begin{equation}
\tau(T)\propto\lbrack\delta\mathcal{A}(T)+2n_{T}(T)]^{-1}\text{ \ ,}%
\label{TauTdep}%
\end{equation}
where $\delta$ is a constant that depends only on the photoexcitation
intensity. In a narrow band semiconductor $n_{T}$ depends on the shape of the
DOS in the energy range $\varepsilon\approx T$ around the chemical potential
(here and further $k_{B}=1$). Generally, the number density of thermally
excited EHPs across the gap (in this case indirect hybridization gap, $E_{hg}%
$) is given by
\begin{equation}
n_{T}\simeq T^{p}\exp(-E_{hg}/2T)\text{ ,}%
\end{equation}
where $p$ is of the order of 1 (0.5 for a BCS superconductor), depending on
the shape of the DOS near the gap edge. Neither the size of $E_{hg}$, the exact shape of the low
energy DOS (anisotropy, possible impurity levels within the gap), or 
the $T$ dependence of $E_{hg}$ is well known in intermediate
valence systems, therefore there is some ambiguity in determining $E_{hg}$.
However, as the main $T$ dependence in $n_{T}$ comes from the exponential
term, a rough estimate of the size of $E_{hg}$ can be obtained. For $T>50$ K a
good agreement of the data with this simple model can be obtained. Since
$\tau$ was found to be independent on $F$ over the range of $F$ studied
\cite{comment2} we can assume that $\tau(T)\approx n_{T}^{-1}$. Indeed Fig. 2
shows that there is a good agreement between $\tau(T)$ and $n_{T}%
^{-1}(T)\propto(\mathcal{A}^{-1}-1)^{-1}$. Fit with $n_{T}^{-1}\propto
T^{-0.5}\exp(E_{hg}/2T)$ gives a values for the indirect HG of $E_{hg}%
\approx16$ meV. This value is lower than the values of the pseudogap obtained
from the inelastic neutron scattering data (30 meV),\cite{Osborn} as well as
the value inferred from the optical conductivity data on YbAl$_{3}$ (60
meV).\cite{Okamura} The relationship between the observed pseudogaps in the
spin (neutrons) and charge (optical) excitation spectra is unclear. On the
other hand, the interpretation of the optical conductivity data is also not
straightforward,\cite{Okamura} since the indirect transitions are forbidden by
the momentum conservation law. Furthermore, the relaxation dynamics bottleneck 
is expected to be governed by the gap minimum, therefore taking the uncertainties 
in the interpretation of both experimental results the agreement is reasonable.
Given the fact, that the mid infrared peak at
0.25 eV,\cite{Okamura} which is in the intermediate valence systems commonly
interpreted as the direct optical transition across the hybridization
gap,\cite{Degiorgi} is in YbAl$_{3}$ present way above room temperature,\cite{Okamura} 
it is expected that the same is true also for the indirect gap, as implied by the present study.

\begin{figure}[h]
\begin{center}
\includegraphics[clip=true,width=0.85\linewidth]{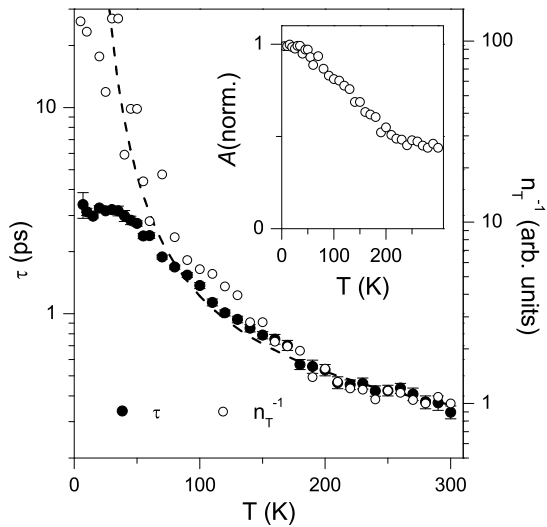}
\end{center}
\caption{T dependence of $\tau$ and the inverse density of thermally excited
quasiparticles, $n_{T}^{-1}$, for YbAl$_{3}$. $n_{T}^{-1}$ was determined from
the T dependence of the normalized amplitude of the photoinduced reflectivity
change, $\mathcal{A}$ (inset), via $n_{T}\propto\mathcal{A}^{-1}-1$. The data
are fit with $n_{T}\simeq T^{1/2}\exp(-E_{hg}/2T)$, where $E_{hg}=16$ meV is
the size of the indirect HG.}%
\end{figure}

Figure 3 presents the $T$ dependence of $\tau$, obtained by the single
exponential fit, on a series of Yb$_{1-x}$Lu$_{x}$Al$_{3}$ samples. Over a
large range of doping, $0\leq x\leq0.3$, all $\tau(T>T^{\ast})$ fall almost on the
same curve. Since $\tau(T>T^{\ast})$ can be well understood in terms of the RT
scenario,\cite{HE2006,Review2006} the results imply that the HG persists over
wide range of doping. We should note that there is some variation in the gap 
extracted from the fit to the relaxation times of individual doping levels 
(10 meV $<E_{hg}<$ 21 meV) - see the two fits in Figure 3. However, there is no systematic dependence of 
$E_{hg}(x)$. At $x=0.4$, however, a qualitative change in the relaxation dynamics 
is observed, with relaxation showing a very weak temperature dependence, much like 
in the metallic LuAl$_{3}$. This observation can be attributed to the loss of the long 
range order at $x=0.4$, consistent with the inelastic neutron scattering data.\cite{Osborn}
Based on the above we argue that the hybridization gap in Yb$_{1-x}$Lu$_{x}$Al$_{3}$ persists,
nearly unchanged, up to $x\simeq 0.3$, with the loss of the long range order to appear only near $x\simeq 0.4$.  

\begin{figure}[th]
\begin{center}
\includegraphics[clip=true,width=0.85\linewidth]{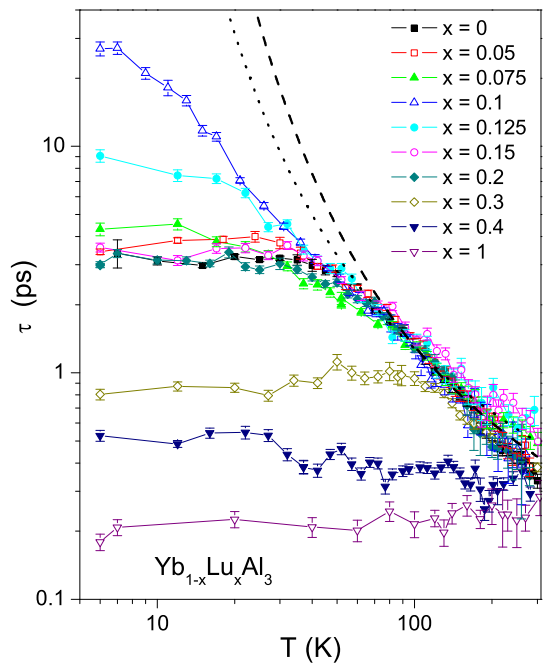}
\end{center}
\caption{(color online) The temperature dependence of $\tau$, extracted using
a single exponential decay fit, for Yb$_{1-x}$Lu$_{x}$Al$_{3}$ series. The
dashed and dotted lines present fits with $\tau\propto T^{-0.5}\exp
(E_{hg}/2T)$ with $E_{hg}=17$ meV and $E_{hg}=10$ meV, respectively.}%
\end{figure}

Below $T^{\ast}(x)$, however, $\tau$ saturates at $\tau_{\ast}(x)$. Generally,
the saturation of $\tau$ at low $T$ is expected within the RT model when
$n_{T}(T)$ becomes smaller than $\delta\mathcal{A}(T)\ \ $- see
Eq.(\ref{TauTdep}). However in this case the low temperature relaxation time,
$\tau_{\ast}$, should depend on the excitation fluence. The fact that
relaxation dynamics in this low perturbation experiment is found to be
independent of excitation intensity effectively rules out this
interpretation.\cite{comment2} Alternatively, we can assume (quite generally),
that the relaxation of the excited state proceeds via two independent
channels. In the case of two competing relaxation processes the recovery
dynamics will always be governed by the fastest of the two. If one of the
processes is slow and temperature independent, while the other shows strong
temperature dependence, becoming slower and slower as the temperature is
reduced, one would expect to observe a crossover from the temperature
dependent to a temperature independent relaxation at the temperature where the
two timescales become comparable. Clearly, this scenario is consistent with
the experimentally observed temperature dependence of the relaxation time.
While at temperatures above $T^{\ast}$ the temperature dependence of both
amplitude and relaxation time are in very good agreement with the relaxation
bottleneck model - see Fig. (2), where the bottleneck in relaxation is a
result of the presence of the hybridization gap near the Fermi level, the
question arises which temperature independent relaxation process could lead to
the observed crossover into T-independent relaxation below $T^{\ast}$?

\section{Analysis and Discussion}

Measurements of the de Haas-van Alphen (dHvA) effect in YbAl$_{3}$ and
LuAl$_{3}$ indicate a large mean free path, $l\approx120-150$
nm,\cite{dhvaybal,dhvalual} several times larger than the optical penetration
depth $\lambda$ ($\lambda_{YbAl_{3}}=22$ nm and $\lambda_{LuAl_{3}}=17$
nm).\cite{OptCond} From the T dependence of resistivity (inset to Fig. 5) it
follows that $l>\lambda$ up to $\approx50$ K, well into experimentally
accessible $T$ range. Hence one can expect that in addition to the recombination
process with characteristic time $\tau$ a ballistic transport of hot electrons
\cite{Wolf} will take place on a timescale of $\lambda/v_{F}$, where $v_{F}$
is the Fermi velocity. From the dHvA data \cite{dhvaybal,dhvalual} one obtains
$v_{F}\approx4$ $10^{4}$ ($1$ $10^{6}$) m/s in YbAl$_{3}$(LuAl$_{3}$) giving
the characteristic timescale for the ballistic transport out of the probed volume
of $\lambda/v_{F}=0.5$ ps and $20$ fs, respectively. Since $v_{F}$ is a weak
function of temperature, the competition between the strongly $T$ dependent
recombination across the HG and a T-independent ballistic transport can
account for the observed temperature dependence of the relaxation time in
Yb$_{1-x}$Lu$_{x}$Al$_{3}$.

To describe the relaxation process for such a case we use the Boltzmann
kinetic equation with the collision integral in the $\tau$%
-approximation,\cite{Future}
\begin{equation}
{\frac{\partial f}{{\partial t}}}+v_{F}\cos{(\theta)}{\frac{\partial
f}{{\partial x}}}=-{\frac{f}{{\tau(\epsilon,\theta)}}}.\label{Eq1}%
\end{equation}
Here $f(t,x,\epsilon,\theta)$ is the non-equilibrium correction to the
equilibrium distribution function, which depends on time $t$, distance from
the surface $x$, energy $\epsilon$ relative to the Fermi energy, and the angle
$\theta$ between the velocity and the transport direction, $x$. The relaxation
time $\tau$ in general depends on the energy and the angle. Here we assume
that the particle-hole symmetry is preserved, so that there is no electric
field. Also, if the Fermi energy is large compared to the photon energy, one
can neglect the dependence of the speed of hot electrons and holes on their
relative energy, i.e. $v\approx v_{F}$. Eq.(\ref{Eq1}), supplemented by the
initial condition $f(t=0,x,\epsilon,\theta)=F(\epsilon,x)$, describing the
initial distribution of the hot quasiparticles after photoexcitation, has the
solution:
\begin{equation}
f(t,x,\epsilon,\theta)=\exp{(-t/\tau(\epsilon,\theta))}F(\epsilon,x-v_{F}%
\cos{(\theta,}t)).\label{f}%
\end{equation}
The spatial and time distribution of the electron-hole pair density is found
as
\begin{equation}
n(x,t)=\int_{0}^{\infty}d\epsilon\int_{0}^{\pi}f(t,x,\epsilon,\theta
)\sin{(\theta)}d\theta.\label{n}%
\end{equation}
For comparison with the experimental data we assume that $\tau$ is energy and
angle independent, $\tau(\epsilon,\theta)=\tau$ and take the Gaussian form of
the energy integrated excitation profile at $t=0$, $\int_{0}^{\infty
}F(\epsilon,x)d\epsilon={\frac{2}{\sqrt{\pi}}}\exp{(-x^{2}/\lambda^{2})}%
$.\cite{Gaussian} $\Delta R/R$ is proportional to the number of photoinduced
carriers within ${\lambda}$. Integrating Eq.(\ref{n}) with the Gaussian probe
profile, $\exp{(-x^{2}/\lambda^{2})}$,\cite{Gaussian} and normalizing to $1$
at $t=0$ one obtains
\begin{equation}
\Delta R/R\propto n(t)=\sqrt{\frac{\pi}{{2}}}{\frac{\lambda}{{v_{F}t}}%
}e^{-t/\tau}erf(v_{F}t/\sqrt{2}\lambda).\label{Ballistic}%
\end{equation}
Eq.(\ref{Ballistic}) has two regimes. On a short time scale, $t\leq
\lambda/v_{F}$, it reduces to
\begin{equation}
n(t)=\exp{(-t/\tau-v_{F}^{2}t^{2}/2\lambda^{2})}\text{ ,}%
\end{equation}
while for $t\gg\lambda/v_{F}$ we have%
\begin{equation}
n(t)={\frac{\lambda}{{v_{F}t}}}e^{-t/\tau}\text{.}%
\end{equation}

\begin{figure}[th]
\begin{center}
\includegraphics[clip=true,width=0.85\linewidth]{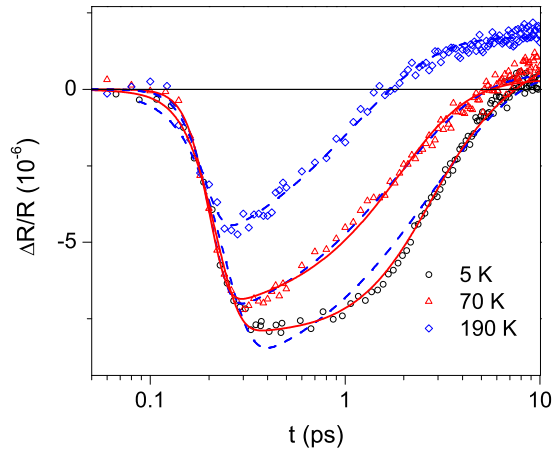}
\end{center}
\caption{(color online) The data recorded on YbAl$_{3}$ at selected
temperatures fit with single exponential decay (dashed blue line) and with
Eq.(\ref{Ballistic}) (solid red line), where $\lambda/v_{F}=1.33$ ps. At low
temperatures the fit is entirely governed by the $\lambda/v_{F}$ term (with $\tau>
5$ ps), while above 50 K the relaxation is best fit with a single exponential
decay.}%
\end{figure}

Figure 4 presents the photoinduced reflectivity traces in YbAl$_{3}$ at a few
selected temperatures. We fit the photoinduced reflectivity traces with
Eq.(\ref{Ballistic}) - solid lines and compare it to the single exponential
decay (dashed).

At $T<T^{\ast}$ the model clearly describes the experimental data much better
than a single exponential decay. In fact, the fit is mainly governed by the
$\lambda/v_{F}$ ratio (best fit gives $\lambda/v_{F}=1.33\pm0.05$ ps in good
agreement with the above estimate) while ${\tau}$ can be anywhere between
5-100 ps. The situation is reversed at $T>T^{\ast}$, where the fit is entirely
governed by ${\tau}$. Since the mean free path quickly decreases below
$\lambda$ upon increasing T, e.g. in YbAl$_{3}$ this should happen at
$\approx50$ K (see inset to Fig. 5), the ballistic transport becomes
ineffective. Correspondingly, the fit to the 70 K data in Fig. 4 with fixed
$\lambda/v_{F}=1.33$ clearly becomes inadequate. In fact, the data are much
better fit with the single exponential decay (dashed blue line). Thus,
$T^{\ast}$ can be thought of as the temperature where $l\approx\lambda$; below
$T^{\ast}$ the relaxation is dominated by the ballistic transport, while above
this temperature relaxation is governed by the recombination across the
indirect HG. In earlier studies \cite{Review2006} such a saturation of $\tau$
at low $T$ was not observed. This can be understood, since at experimentally
accessible temperatures of $T\gtrsim10$ K the mean free path in YbAl$_{3}$ is
by far the largest among the systems studied. Still, comparison of dynamics in
YbCdCu$_{4}$ and YbAgCu$_{4}$ (Fig. 5 of Ref. [4]) shows an onset of
saturation at $\approx20$ K in YbAgCu$_{4}$. This feature is absent in
YbCdCu$_{4}$. Since the mean free path in YbCdCu$_{4}$ is considerably smaller
than in YbAgCu$_{4}$,\cite{Sarrao} this observation is consistent with the
proposed scenario.

\begin{figure}[th]
\begin{center}
\includegraphics[clip=true,width=0.85\linewidth] {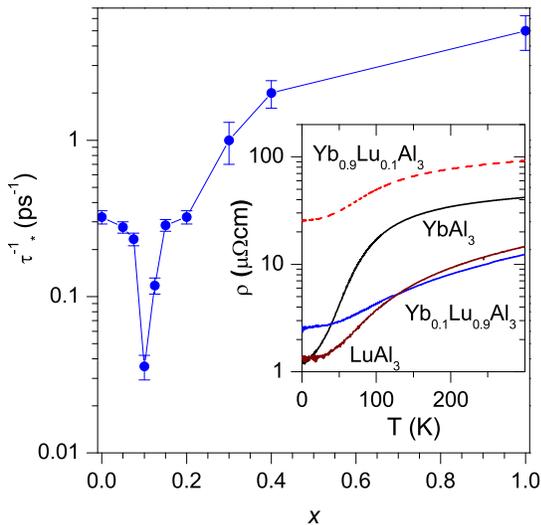}
\end{center}
\caption{(color online) Doping dependence of $\tau_{\ast}^{-1}$ for Yb$_{1-x}%
$Lu$_{x}$Al$_{3}$. Inset shows the $T$ dependence of resistivity for several
doping levels.}%
\end{figure}

The doping dependence of the low temperature relaxation rate, $\tau_{\ast
}^{-1}(x)$, is plotted in Fig. 5. Initially $\tau_{\ast}^{-1}$ shows a weak
decrease from $\tau_{\ast}^{-1}(x=0)\approx0.2$ ps$^{-1}$, followed by a rapid
drop at $x=0.1$. Further increase in $x$ results in an increase in $\tau
_{\ast}^{-1}$ up to $\approx$ 5 ps$^{-1}$. Within the above scenario,
following the $\lambda/v_{F}$ ratio, one would expect $\tau_{\ast}^{-1}%
$(LuAl$_{3}$) $\approx20$ $\tau_{\ast}^{-1}$(YbAl$_{3}$), as observed. The
strong drop in $\tau_{\ast}^{-1}$ near $x=0.1$ can be understood as being a
result of the decrease of $l$ below $\lambda$, consistent with the resistivity
data (see insert to Fig. 5). Indeed, in Yb$_{0.9}$Lu$_{0.1}$Al$_{3}$ the
expected low $T$ divergence of $\tau$ due to the relaxation bottleneck
\cite{HE2006,Review2006} is largely recovered (see Fig.3). Thus, the
qualitative doping dependence of $\tau_{\ast}^{-1}$ in Yb$_{1-x}$Lu$_{x}%
$Al$_{3}$ can be understood as a competition between decreasing $\lambda
/v_{F}$ ratio and doping induced disorder. Still, the underlying origin of the
very sharp anomaly at $x\approx0.1$ is unclear. Noteworthy, this doping level
corresponds well with the doping level where low temperature coherence effects
are strongly suppressed.\cite{Cornelius,Bauer04}

\section{Conclusions}

We studied the carrier relaxation dynamics in Yb$_{1-x}$Lu$_{x}$Al$_{3}$, an
intermediate valence heavy fermion system where doping was continuously tuned
between a heavy fermion and normal metal. In YbAl$_{3}$ there seems to be an
overall agreement that its low temperature properties are governed by a
hybridization of conduction electrons with spatially extended wave functions
and localized f orbitals, which can be described by the Anderson lattice
model. Since photoexcited carrier relaxation was found to be very sensitive to
the appearance of the gap in the charge excitation spectrum, the aim of this
work was to utilize time-resolved optical spectroscopy to determine the doping
level at which long range order, leading to the indirect hybridization gap in
intermediate valence heavy electron systems, is suppressed. By performing
systematic studies of photoexcited carrier relaxation dynamics in Yb$_{1-x}%
$Lu$_{x}$Al$_{3}$ system, where doping was continuously varied between $0\leq
x\leq1$, we show that the hybridization gap persists up to the critical doping
$x\approx0.4$. This is consistent with inelastic neutron scattering
data.\cite{Osborn} Indeed, the percolation threshold for the cubic system is
$\approx0.3$.\cite{Ziff} This observation implies that the Kondo lattice is a
very robust feature, and suggest the Anderson lattice model to be a minimum
model for the description of this class of materials.

In Yb$_{1-x}$Lu$_{x}$Al$_{3}$, below some doping dependent temperature
$T^{\ast}(x)$, the relaxation time divergence due to the relaxation bottleneck
is cut. This observation can be naturally explained by considering a parallel
temperature independent relaxation channel. Since YbAl$_{3}$ is known for its
large electronic mean free path, we consider ballistic electronic transport
out of the probed volume as a possible competing process. We developed a
theoretical model, describing the dynamics of excess quasiparticle density for
such a case. Comparison with the experimental data showed that the functional
form of relaxation below $T^{\ast}(x)$, the temperature dependence of relaxation 
time, as well as the doping dependence of the low temperature relaxation rate 
in Yb$_{1-x}$Lu$_{x}$Al$_{3}$ can be well accounted for by this model with the 
dominant ballistic electron transport at low $T$. The strikingly sharp anomaly 
in $\tau_{\ast}^{-1}(x)$ at $x\approx0.1$ clearly requires further experimental 
and theoretical work.

\begin{acknowledgments}
This work was supported by Sofja-Kovalevskaja Award from the Alexander von
Humboldt Foundation, Zukunftskolleg and Center for Applied Photonics at the
Uni. Konstanz, the Laboratory Directed Research and Development program at Los
Alamos National Laboratory, and Center for Integrated Nanotechnologies at LANL.
\end{acknowledgments}


\begin{thebibliography}{99}
\bibitem {comment2}Fluence was kept low to avoid sample heating. Over the
fluence range $F=0.1-2$ $\mu$J/cm$^{2}$ no nonlinearities were observed, i.e.
the photoinduced transient scales with $F$.
\bibitem {Gaussian}Such choice of the excitation profile is used to satisfy
the condition $\partial F/\partial x=0$ at $x=0$. This condition means that
quasiparticles cannot escape from the sample through the surface. The results
are independent of the particular choice of excitation profile.
\end{thebibliography}
\end{document}